\documentclass[a4paper, twocolumn, amsfonts, amssymb, amsmath, aps, reprint, showkeys, nofootinbib]{revtex4-2}
\usepackage[english]{babel}
\usepackage[utf8]{inputenc}
\usepackage[colorinlistoftodos, color=green!40, prependcaption]{todonotes}
\usepackage{amsthm}
\usepackage{mathtools}
\usepackage{physics}
\usepackage{xcolor}
\usepackage{graphicx}
\usepackage{adjustbox}
\usepackage{placeins}
\usepackage[T1]{fontenc}
\usepackage{csquotes}
\usepackage{bbm}
\usepackage{enumerate}
\usepackage{color}
\usepackage{latexsym}
\usepackage{times,txfonts}
\usepackage{microtype}
\usepackage{parskip}
\usepackage[pdftex, pdftitle={Article}, pdfauthor={Author}]{hyperref} 
\begin{document}
\title{Reconciling nonlinear dissipation with the bilinear model of two Brownian particles}

\author{Elisa I. Goettems}
    \email {elisagtt@ifsc.usp.br}
    \affiliation{Instituto de F\'isica de S\~ao Carlos, Universidade de S\~ao Paulo, CP 369, 13560-970, S\~ao Carlos, SP, Brazil}
\author{Ricardo J. S. Afonso}
    \email {ricardo.afonso@ifsc.usp.br}
   \affiliation{Instituto de F\'isica de S\~ao Carlos, Universidade de S\~ao Paulo, CP 369, 13560-970, S\~ao Carlos, SP, Brazil}
\author{Diogo O. Soares-Pinto}
    \affiliation{Instituto de F\'isica de S\~ao Carlos, Universidade de S\~ao Paulo, CP 369, 13560-970, S\~ao Carlos, SP, Brazil}
\author{Daniel Valente}
    \affiliation{Instituto de F\'isica, Universidade Federal de Mato Grosso, 78060-900 Cuiab\'a MT, Brazil}

\date{\today} 

\begin{abstract}
The Brownian motion of a single particle is a paradigmatic model of the nonequilibrium dynamics of dissipative systems.
In the system-plus-reservoir approach, one can derive the particle's equations of motion from the reversible dynamics of the system coupled to a bath of oscillators representing its thermal environment. However, extending the system-plus-reservoir approach to multiple particles in a collective environment is not straightforward, and conflicting models have been proposed to that end. Here, we set out to reconcile some aspects of the nonlinear and the bilinear models of two Brownian particles.
We show how the nonlinear dissipation originally derived from exponential system-reservoir couplings can alternatively be obtained from the bilinear Lagrangian, with a modified spectral function that explicitly depends on the distance between the particles. As applications, we discuss how to avoid the anomalous diffusion from the standard nonlinear model, as well as how to phenomenologically model a hydrodynamic interaction between a pair of Brownian particles in a viscous fluid.

\end{abstract}


\maketitle

\section{Introduction}\label{sec:introduction}

The Brownian motion is a representative example of nonequilibrium dissipative dynamics.
Theoretically, the irreversible dynamics of a subsystem can emerge from a reversible dynamics of the global system. 
To achieve that, one way is to phenomenologically model the particle's environment as a set of independent oscillators, each linearly coupled to the system of interest.
This is the so called system-plus-reservoir approach.
By tuning the spectral function, which determines the weight of each frequency mode on the particle's dissipation rate, one can recover experimental observations.
This framework has proven useful across classical and quantum domains
~\citep[][]{zwanzig2001nonequilibrium,weiss2012quantum,caldeira2014introduction}.

Let us consider two particles immersed in the same bath. This can be relevant, for instance, when environment-induced effects on multiple degrees of freedom are being investigated, as in the cases of biologically-inspired problems ~\citep[][]{gelin2011,valente21}, of non-Markovianity ~\citep[][]{fred}, of synchronization ~\citep[][]{henriet} and of quantum entanglement ~\citep[][]{duarte2009effective,paz2008dynamics,zell,valente10}, to name a few.
One way to address this problem is to employ the so called bilinear model, which assumes that each particle is linearly coupled to the same set of oscillators ~\citep[][]{paz2008dynamics,miled,zell,gelin2011,henriet,fred}.
However, such a bilinear coupling may lead to unphysical results, namely, the free-particle motion of the relative coordinate, and the absence of mutual effects between proximal particles, as pointed out by Duarte and Caldeira ~\citep[][]{duarte2006effective}.
To solve these issues, a nonlinear model for the system-environment couplings has been devised, which not only recovers the well-known single-particle case, but also predicts dissipation rates that are nonlinear functions of the distance between the pair of Brownian particles ~\citep[][]{duarte2006effective}.

Despite the successes of the nonlinear model for two Brownian particles, the bilinear approach has its merits.
Experimentally, it can yield correct results, as in the study of heat flow between Brownian particles ~\citep[][]{berut2014energy}, and in a recent demonstration of environment-induced entanglement in the optical domain ~\citep[][]{davidovich18}.
Theoretically, linearity allows the model to be exactly solvable in the quantum and the classical regimes, making it a desirable tool.
Additionally, the original spirit of the system-plus-reservoir approach as a search for a simplified description of an otherwise untractable interaction with a complex environment sounds more in line with the approach of the bilinear model.

Here, we address the following question. 
Is there an alternative way to avoid the shortcomings of the standard bilinear model, without recurring to nonlinear system-environment couplings?
To that end, we introduce a physically motivated spectral function that explicitly depends on the relative distance between a pair of Brownian particles.
By doing so, we obtain, from the bilinear model, the nonlinear dissipation that had been originally derived from exponential system-environment couplings ~\citep[][]{duarte2006effective}.
Our method thereby reveals a simpler and more versatile way to phenomenologically model nonlinear environment-induced forces.
In particular, we discuss how to avoid an anomalous diffusion found in the nonlinear model, and also how our theory can describe hydrodynamic correlations of multiple Brownian particles
~\citep[][]{ermak1978brownian}.

In Sec.\ref{sec:nonlinear}, we revisit the bilinear and nonlinear models.
In Sec.\ref{sec:dynamics}, we show our main result, namely, how nonlinear dissipation can be obtained from the bilinear Lagrangian.
For that, we define a spectral function that depends on the relative distance between the particles (Sec.\ref{mainresult}).
We show how our modified spectral function enables us to bridge the dissipation rates from the standard nonlinear and the standard bilinear models (Sec.\ref{sec:recovering}).
We discuss how to avoid the anomalous diffusion of the original nonlinear model (Sec.\ref{sec:anomalousdiffusion}), and how our approach can be applied to model a pair of Brownian particles sharing a hydrodynamic environment (Sec.\ref{sec:hydrodynamics}).
Finally, in Sec.\ref{sec:conclusions} we present our conclusions and further considerations. 
Detailed derivations are presented in the Appendix\ref{appendix1:apendix1}.

\section{Preliminary Remarks}\label{sec:nonlinear}

\subsection{Standard bilinear model}

In the bilinear model ~\citep[][]{paz2008dynamics}, the Lagrangian of two classical Brownian particles immersed in a collective environment reads 
\begin{align}
    \begin{split}
       L = &\frac{m}{2}(\dot{x}_1^2 + \dot{x}_2^2)-\frac{1}{2}\sum_{k}R_k\left(C_k^{(1)}x_1+C_k^{(2)}x_2\right)\\
       &-\frac{1}{2}\sum_{k}
       \frac{1}{m_k\omega_k^2}
       \left(C_k^{(1)}x_1+C_k^{(2)}x_2\right)^2
       +\sum_{k}\frac{m_k}{2}(\dot{R}_k^2-\omega_k^2R_k^2).
    \end{split}
\end{align}
Here, $\dot{x}_{1,2}$ and $x_{1,2}$ are the velocity and the positions of the particles with mass $m$. $R_k$ is the position of the $k$-th bath oscillator, with frequency $\omega_k$ and mass $m_k$.
The coupling between the system and each oscillator is assumed to be linear in their positions, with distinct coupling strengths $C_k^{(i)}$.
The counterterm is added so as to offset the environment-induced modification on the external potential.

One can derive the Euler-Lagrange equations for both the system and the bath. From now on, we define the center of mass and relative coordinates as
\begin{align}\label{ref:q}
    \begin{split}
        q=\frac{x_1+x_2}{2},\\
    \end{split}
\end{align}
and
\begin{align}\label{ref:u}
    \begin{split}
        u=x_1-x_2,
    \end{split}
\end{align}
which yield the following equations of motion,
\begin{align}\label{ref:lineareq}
    \begin{split}
        m\ddot{q}(t)+(\eta+\eta_{12})\dot{q}(t)=f_q(t),\\
        m\ddot{u}(t)+(\eta-\eta_{12})\dot{u}(t)=f_u(t).
    \end{split}
\end{align}
One can interpret $f_q(t)$ and $f_u(t)$ as the fluctuating forces for the center of mass and the relative coordinate,
\begin{align}
    \begin{split}
        f_q(t) &= -\sum_k \frac{(C_k^{(1)}+C_k^{(2)})}{2}\\
        &\times \left(\dot{R}_k(0)\frac{\sin{\omega_k t}}{\omega_k}+\Tilde{R}_k(0)\cos{\omega_k t}\right),
    \end{split}
\end{align}
\begin{align}
    \begin{split}
        f_u(t) &= \sum_k (C_k^{(1)}-C_k^{(2)})\\
        &\times \left(\dot{R}_k(0)\frac{\sin{\omega_k t}}{\omega_k}+\Tilde{R}_k(0)\cos{\omega_k t}\right).
    \end{split}
    \label{fubilinear}
\end{align}
where $\Tilde{R}_k(0) = R_k(0)+[C_k^{(i)}x_i(0)+C_k^{(j)}x_j(0)]/m_k \omega_k^2$. The statistical properties of these forces stem from the initial state of the total system.

For the dissipation term, we assume an Ohmic bath ~\citep[][]{caldeira2014introduction}, so the usual spectral functions read
\begin{align}
    J_i(\omega) = \frac{\pi}{2}\sum_k \frac{C_k^{(i)2}}{m_k\omega_k}\delta(\omega - \omega_k )\equiv \eta \omega \Theta (\Omega - \omega),
\end{align}
with the high frequency cutoff $\Omega$ ~\citep[][]{guinea1984friction, hedegaard1987quantum}, and with $\Theta$ being the Heaviside step function.
Similarly, a mixed spectral function appears,
\begin{align}
J_{12}(\omega) = \frac{\pi}{2}\sum_k \frac{C_k^{(1)}C_k^{(2)}}{m_k \omega_k}\delta(\omega-\omega_k) \equiv \eta_{12}\omega\Theta(\Omega-\omega),    
\label{eta12}
\end{align}
as an indication of bath-mediated interactions between the particles. 
This introduces the dissipation rates $\eta$ and $\eta_{12}$.

Two properties of this model call our attention.
First, the case of identical couplings, $C_k^{(1)} = C_k^{(2)}$, (a reasonable hypothesis, as far as two proximal particles in the same environment are concerned) implies that
$\eta = \eta_{12}$ and $f_u(t) = 0$, leading to vanishing dissipation and fluctuating forces (see Eqs.(\ref{ref:lineareq}) and (\ref{fubilinear})).
It means that a free-particle motion is found, namely,
\begin{align}
    \ddot{u}(t) = 0.
\end{align}
Second, even for nonidentical couplings one finds that $\eta_{12}$ is independent of the distance between the particles. 
Both considerations suggest instantaneous effects between spatially separate entities.
As discussed by Duarte and Caldeira ~\citep[][]{duarte2006effective}, these are undesirable features, arising from the lack of an appropriate length scale for environment-mediated phenomena.

\subsection{Standard nonlinear model}
\label{sec:nonlinearmodel}
In the nonlinear model as introduced by Duarte and Caldeira in Ref.~\citep[][]{duarte2006effective}, the Lagrangian reads 
\begin{align}
    \begin{split}
       &L = \frac{m}{2}(\dot{x}_1^2 + \dot{x}_2^2) + \frac{1}{2}\sum_{k}m_k\left(\dot{R}_k\dot{R}_{-k}-\omega_k^2R_k R_{-k}\right)\\
       &-\frac{1}{2}\sum_{k}\left[(C_{-k}(x_1)+C_{-k}(x_2))R_k+(C_k(x_1)+C_k(x_2))R_{-k}\right],
    \end{split}
\end{align}
where 
\begin{align}
C_k(x) =\kappa_ke^{ikx}.   
\end{align}
The exponential couplings between the bath and the particles guarantee homogeneity and translational invariance.
Note that the index $k$ now has dimensions of $[L]^{-1}$, explaining why it introduces the required length scale.

The equations of motion are
\begin{align}
    \begin{split}
        m\ddot{x}_i(t)&+\int_0^t\mathbf{K}(x_i(t)-x_i(t'),t-t')\dot{x}_i(t')dt'\\
        &+\int_0^t\mathbf{K}(x_i(t)-x_j(t'),t-t')\dot{x}_j(t')dt'\\
        &+\dfrac{\partial }{\partial x_i}V(x_i(t)-x_j(t))=F_i(t)
    \end{split}
\end{align}
with $i,j = 1,2$, and once again $F_i(t)$ can be interpreted as the fluctuating force.
The dissipation kernels are 
\begin{align}
    \mathbf{K}(r,\tau) &=\sum_k\int_{0}^{\infty} d\omega 2k^2\kappa_k\kappa_{-k}  \nonumber \\
   &\times \frac{\mathrm{Im} \chi_k^{(i)}(\omega)}{\pi \omega} \cos{\omega_k \tau} \cos{kr},
\end{align}
written in terms of the imaginary part of the dynamical response of the environment oscillators, namely,
\begin{align}
 \mathrm{Im} \chi_k^{(i)}(\omega) \equiv\frac{\pi}{2 m_k\omega_k}\delta(\omega-\omega_k).   
\label{defchi}
\end{align}
In the present model, $\mathrm{Im} \chi_k^{(i)}(\omega)$ is equivalent to the spectral function in that it allows for the transformation of a discrete set of oscillators into a continuum.
The delta function is thus replaced by a Lorentzian peaked around $\omega_k$.
The next step is to focus on the low-frequency limit of that Lorentzian so as to recover the Ohmic regime (linear in $\omega$).
This justifies the approximation
\begin{align}\label{eq:suscep_i}
    \mathrm{Im} \chi_k^{(i)}(\omega) \approx \Tilde{f}(k)\omega \Theta(\Omega-\omega).
\end{align}
Nonlinear equations of motion are obtained, such that
\begin{align}
    m\ddot{q}(t)+\left(\eta+\eta_e[u(t)]\right)\dot{q}(t) = f_q(t),
    \label{ref:nonlineareq1}
\end{align}
and
\begin{align}
   m\ddot{u}(t)+\left(\eta-\eta_e[u(t)]\right)\dot{u}(t)+V_e'(u(t)) = f_u(t).
\label{ref:nonlineareq2}
\end{align}
Here, 
\begin{align}
    V_e(u) \equiv -\frac{2\Omega \eta}{\pi k_0^2(k_0^2u^2+1)}
\label{ve}
\end{align}
represents an environment-induced potential, which depends on the relative distance.
Similarly,
\begin{align}
    \eta_e[u]\equiv\eta \frac{(1-3k_0^2u^2)}{(k_0^2u^2+1)^3}
\label{etae}
\end{align}
describes a distance-dependent bath-mediated dissipation rate.
The constant $k_0$ is a characteristic inverse length introduced when the summation over $k$ is transformed into an integral, $\sum_k \rightarrow (L/2\pi) \int dk$, and a density of spatial modes is postulated,
\begin{align}
    \eta g(k) \equiv \frac{L}{2 \pi}\kappa_k\kappa_{-k}\Tilde{f}(k),
\label{continuumnonlin}
\end{align}
such that $\int_0^\infty dk g(k)k^2 = 1$.
Finally, the choice for
\begin{align}
    g(k) = \left(\frac{1}{2k_0^3}\right) e^{-k/k_0}
\end{align}
explains how $k_0$ is defined in Ref.~\citep[][]{duarte2006effective}.

Equations (\ref{ref:nonlineareq1}) to (\ref{etae}) describe rich environment-induced behaviors for proximal particles, while recovering the independent Brownian movement for arbitrarily large distances.
The free-particle anomaly found in the bilinear model is no longer present, since both the dissipative and the fluctuating forces are finite for any finite $u(t)$.
The statistical properties of the fluctuating forces are such that (see Appendix \ref{appendix1:apendix1})
\begin{align*}
    \langle{f_q(t)}\rangle&=\langle{f_u(t)}\rangle=0,\\  \langle{f_q(t)f_q(t')}\rangle&= k_BT(\eta+\eta_e[u(t)]) \delta(t-t'),\\
    \langle{f_u(t)f_u(t')}\rangle&= 4k_BT(\eta-\eta_e[u(t)]) \delta(t-t'),
\end{align*}
in agreement with the fluctuation-dissipation theorem.

\section{Results}\label{sec:dynamics}

\subsection{Nonlinear dissipation from a bilinear Lagrangian: modified spectral function}
\label{mainresult}

In this section we show how the nonlinear dissipation term $\eta_e[u(t)]$ from Eqs.(\ref{ref:nonlineareq1}) and (\ref{ref:nonlineareq2}) can be obtained from a bilinear model.
Following Ref.~\citep[][]{paz2008dynamics}, we assume a Lagrangian that breaks translational invariance both in the system degrees of freedom and in the couplings to the bath,
\begin{align}
    \begin{split}
       L &=\frac{m}{2} (\dot{x}_1^2+\dot{x}_2^2)  - V(x_1)-V(x_2) - m \ c_{12}x_1x_2\\
       &+\sum_{k=1}^{N}\left(\frac{m_k\dot{R}_k^2}{2}-\frac{m_k}{2}\omega_k^2R_k^2\right)-
       \sum_{k=1}^{N}(C_k^{(1)}x_1+C_k^{(2)}x_2)R_k.
    \end{split}
\end{align}%
We have maintained the direct coupling between the particles, proportional to $c_{12}$, as done in Ref.\citep[][]{paz2008dynamics}.
Note that, in order to avoid the anomalous free-particle motion for the relative coordinate ($\ddot{u} = 0$), we have to assume that each particle has a distinct coupling parameter ($C_k^{(1)} \neq C_k^{(2)}$), otherwise the bath decouples from $u = x_1 - x_2$. 
Using similar techniques as in the previous sections, one gets to the following equations of motion for each particle
\begin{align}
    \begin{split}
        m\ddot{x}_i+&\dfrac{dV(x_i)}{dx_i}+m c_{12}x_j\\ &+\sum_{k}\frac{C_k^{(i)2}}{m_k \omega_k^2}\int_0^t\cos{\omega_k(t-t')}\dot{x}_i(t')dt'\\
        &+\sum_{k}\frac{C_k^{(i)}C_k^{(j)}}{m_k \omega_k^2}\int_0^t\cos{\omega_k(t-t')}\dot{x}_j(t')dt'=f_i(t),
    \end{split}
\end{align}
where $i\neq j = 1,2$, and $f_i(t)$ term can be interpreted as the fluctuating force 
\begin{align}
    \begin{split}
        f_i(t) = -\sum_k C_k^{(i)}\left[\dot{R}_k(0)\frac{\sin{\omega_k t}}{\omega_k}+\Tilde{R}_k(0)\cos{\omega_k t}\right],
    \end{split}
\end{align}
where $\Tilde{R}_k(0) = R_k(0)+(C_k^{(i)}x_i(0)+C_k^{(j)}x_j(0))/(m_k \omega_k^2)$.

To evidence the center of mass and the relative coordinate, we rewrite the equations of motion as
\begin{align}\label{centerofmasscoordinate}
    \begin{split}
        &m\ddot{q}(t)+\frac{1}{2}\left(\dfrac{dV(x_1)}{dx_1}+\dfrac{dV(x_2)}{dx_2}\right)+mc_{12}q(t)\\
        &+\frac{1}{2}\int_0^t\left[K_1(t-t')\dot{x}_1(t')+K_2(t-t')\dot{x}_2(t')\right]dt'\\
        &+\int_0^t K_{ij}(t-t')\dot{q}(t')dt'=f_  q(t),
    \end{split}
\end{align}

\begin{align}\label{relativecoordinate}
    \begin{split}
        &m\ddot{u}(t)+\left(\dfrac{dV(x_1)}{dx_1}-\dfrac{dV(x_2)}{dx_2}\right)-mc_{12}u(t)\\
        +&\int_0^t\left[K_1(t-t')\dot{x}_1(t')-K_2(t-t')\dot{x}_2(t')\right]dt'\\
        &-\int_0^tK_{ij}(t-t')\dot{u}(t')dt'=f_u(t).
    \end{split}
\end{align}%

The dissipation kernels in Eqs. (\ref{centerofmasscoordinate}) and  (\ref{relativecoordinate}) are different from the ones in Ref.~\citep[][]{duarte2006effective} in the lack of a spatial dependence of the environment-induced effects, as we show below,
\begin{align}\label{kernel-i}
    &\mathbf{K}_i(t-t')=
    \sum_k2 C_k^{(i)2}
    \int\displaylimits_{0}^{\infty}d\omega 
    \frac{\Im \chi_k^{(i)}(\omega)}{\pi \omega}\cos{\omega(t-t')},\\
    \label{kernel-ij}
    &\mathbf{K}_{ij}(t-t')=\sum_k 2C_k^{(i)}C_k^{(j)}\int\displaylimits_{0}^{\infty}d\omega \frac{\Im\chi_k^{(ij)}(\omega)}{\pi \omega}\cos{\omega(t-t')}.
\end{align}%
We have also used Eq.(\ref{defchi}) to write the kernels in terms of the imaginary part of the bath susceptibility functions.

As far as the single-particle dissipation rates are concerned, we define
\begin{align}
 \eta \equiv \sum_k C_k^{(i)2}\tilde{f}(k),
 \label{etaourmodel}
\end{align}
where $\tilde{f}(k)$ comes from the Ohmic approximation in Eq.(\ref{kernel-i}), i.e., $\Im\chi_k^{(i)}(\omega) \approx \tilde{f}(k)\omega \Theta(\Omega-\omega)$ (similarly to Eq.(\ref{eq:suscep_i})).
To obtain the continuum limit, we replace
$\sum_k C_k^{(i)2}\tilde{f}(k)$ 
by 
$\eta \int dk  \ k^2 g(k)$, 
where
$g(k) = \exp(-k/k_0)/(2k_0^3)$ (as in Sec.\ref{sec:nonlinearmodel}).

The key step in our derivation concerns the two-particle susceptibility $\chi_k^{(ij)}$.
We consider that, when two Brownian particles are sufficiently close, the environment acting on each particle is composite: It is jointly formed by the free environment dynamics plus the perturbation of the other particle dynamics on that environment.
Put differently, each particle ``sees'' an effective bath ``dressed'' by the state of the other one, when these are close enough.
As a consequence, our insight translates into a response function that should depend on the distance between the particles or, more generally, on the relative coordinate $u$.
We thus postulate that, in our model,
\begin{align}
    \Im\chi_k^{(ij)}(\omega) \equiv \Im\chi_k^{(ij)}(\omega,u).
\end{align}
We obtain the Ohmic regime by choosing a linear function in $\omega$, namely,
\begin{align}
    \label{susceptibility-ij}
    \Im\chi_k^{(ij)}(\omega,u)  &\approx h(k,u)\omega \Theta(\Omega-\omega),
\end{align}
where $h(k,u)$ is to be defined.
The explicit choice for $h(k,u)$ will allow us to define the bath-related length scale.
Because our main goal is to reobtain the nonlinear dissipation from the nonlinear model, we write $h(k,u)$ in the form
\begin{align}
    h(k,u)= \tilde{F}(k)G(k,u).
        \label{hku}
\end{align}
Here, $\tilde{F}(k)$ is analogous to $\tilde{f}(k)$ in that it allows us to define
\begin{align}
    \eta_{\mathrm{eff}}[u] \equiv \eta \int dk \  g_{\mathrm{eff}}(k) G(k,u)
    \label{etaeffourmodel}
\end{align}
as the continuum limit of 
$\sum_k C_k^{(i)} C_k^{(j)} \tilde{F}(k) G(k,u)$.
This summation is obtained from applying Eqs. (\ref{susceptibility-ij}) and (\ref{hku}) to (\ref{kernel-ij}).
We have also defined 
\begin{align}
 g_{\mathrm{eff}}(k) \equiv k^2 g(k),
\end{align}
motivated by the fact that we recover the single-particle dissipation rate by choosing $G(k,u) = 1$.

Using Eqs.(\ref{etaourmodel}) and (\ref{etaeffourmodel}), and taking the limit $\Omega \rightarrow \infty$ in Eqs. (\ref{kernel-i}) and (\ref{kernel-ij}), the dissipation kernels become
\begin{align}
    \mathbf{K}_i(t-t')&= 2 \eta \delta(t-t')\\
    \mathbf{K}_{ij}(t-t')&= 2\eta_{\mathrm{eff}}[u]\delta(t-t').
\end{align}
The length scale is now explicit, given by the dependence of the kernel on the relative coordinate $u$.
By tuning $G(k,u)$, one controls the nonlinear dissipation force appearing in the equations of motion, be it in order to recover a certain theoretical model or to explain a specific experiment.

Finally, our equations of motion in the case of free Brownian particles, $V(x_1) = V(x_2) = 0$, read 
\begin{align}\label{eq:motion}
    \begin{split}
        &m\ddot{q}(t)+\left(\eta+\eta_{\mathrm{eff}}[u]\right)\dot{q}(t)=f_q(t)\\
        &m\ddot{u}(t)       +\left(\eta   -\eta_{\mathrm{eff}}[u]\right)\dot{u}(t)=f_u(t),
    \end{split}
\end{align}
where center of mass and relative fluctuating forces are given by $f_q(t) = (f_1+f_2)/2$ and $f_u(t) = f_1-f_2$. 
Note that, although nonlinear dissipation forces have been found, the effective bath-induced potential $V_e[u]$ from Eq.(\ref{ref:nonlineareq2}) could not be recovered in our modified bilinear model.
This suggests that, if a given experiment reveals bath-mediated conservative forces, the theoretical model should probably start from nonlinear system-bath couplings.

\subsection{Recovering the dissipation rates}
\label{sec:recovering}

In order to recover the nonlinear dissipation $\eta_e[u]$ from Eq.(\ref{etae}), we choose 
\begin{align}
G(k,u)=\cos(ku).
\end{align}
We use this in Eq.(\ref{etaeffourmodel}), thus finding that
\begin{align}
    \eta_\mathrm{eff}[u] 
    = \eta \int_0^\infty dk \ k^2 \  \frac{e^{-k/k_0}}{2k_0^3} \  \cos(ku)
    = \eta \frac{(1-3k_0^2u^2)}{(k_0^2u^2+1)^3}.
\label{etaerecovered}
\end{align}
As expected, $\eta_\mathrm{eff}[u] = \eta_e[u]$.

We also recover the constant dissipation rate $\eta_{12}$ shown in Eq.(\ref{eta12}), in the context of the standard bilinear model.
By choosing $G(k,u) = G_0$, we find that
\begin{align}
\eta_{\mathrm{eff}}[u] = \eta G_0 = \eta_{12}.    
\end{align}
This means that our result bridges the standard bilinear and the standard nonlinear models, as far as dissipation is concerned.

\subsection{Avoiding the anomalous diffusion} 
\label{sec:anomalousdiffusion}
It is worth discussing the anomalous diffusion due to the specific form of $\eta_{\mathrm{eff}}[u]$ in Eq.(\ref{etaerecovered}).
This can be seen from the diffusion coefficients appearing in the correlation functions of the Langevin forces, namely,
\begin{align}
 \langle{f_1(t)f_2(t')}\rangle=2D_{12}(u)\delta(t-t'),   
\end{align}
$\langle{f_u(t)f_u(t')}\rangle=2D_{u}(u)\delta(t-t')$,
and
$\langle{f_q(t)f_q(t')}\rangle=2D_{q}(u)\delta(t-t')$.
The connection comes from the fact that
\begin{align}
    D_{12}(u) = \eta_{\mathrm{eff}}[u] k_B T.
    \label{D12}
\end{align}
Similarly, we also find
\begin{align}
 D_u(u) = (\eta - \eta_{\mathrm{eff}}[u]) 2 k_B T,   
\label{Du}
\end{align}
and
\begin{align}
D_q(u) = (\eta + \eta_{\mathrm{eff}}[u]) \frac{k_B T}{2}.
\label{Dq}
\end{align}
The anomalous diffusion arises from
\begin{align}
    \eta_{\mathrm{eff}}[|u|> k_0^{-1}/\sqrt{3}] < 0,
\label{etaneg}
\end{align}
implying a reduction in the diffusion coefficient $D_q$ at intermediate separations, as compared to arbitrarily far apart Brownian particles, as well as an anticorrelation ($D_{12} < 0$) between the Langevin forces acting on the particles.

To avoid the anomalous diffusion, we tune $G(k,u)$.
This allows us to derive a different  behavior that could still fulfill the fluctuation-dissipation relation, while eliminating the anomalous anticorrelation effect.
To be concrete, let us take the example of the spectral functions used to model localized excitons interacting with a bath of acoustic phonons ~\citep[][]{robson,axt}.
Their typical Gaussian features motivate us to set 
\begin{align}
G(k,u) =e^{-\frac{k}{k_0}(k_0 u)^2}.
\label{gauss}
\end{align}
This results in an effective dissipation rate given by
\begin{align}
    \eta_{\mathrm{eff}}[u] = \frac{\eta}{(1+k_0^2u^2)^3}.
    \label{etagauss}
\end{align}
That is, Eq.(\ref{gauss}) guarantees that $\eta_{\mathrm{eff}}[u] \geq 0$, and also recovers the independent Brownian motions of two arbitrarily distant particles (in the limit of $|u|\rightarrow \infty$).
In fact, any positive and convergent function (i.e., $G(k,u)\geq 0$ and $G(k,|u|  \rightarrow \infty) = 0$) is sufficient to guarantee that $\eta_{\mathrm{eff}}[u] \geq 0$, and that $\eta_{\mathrm{eff}}[|u| \rightarrow \infty] = 0$.

\subsection{Hydrodynamics-inspired model} \label{sec:hydrodynamics}

We now consider two Brownian particles immersed in a viscous fluid.
The fluid can mediate interactions between the particles, so that the dissipative forces may depend on the interparticle distance, as shown in Ref.~\citep[][]{ermak1978brownian}.
Here, we restrict our discussion to the limit where the radius of each Brownian particle is vanishingly small as compared to their relative distance (i.e, the regime of validity of the Oseen tensor).
We also assume the one-dimensional limit of the Brownian motion.
In that case, an effective dissipation rate mediated by the hydrodynamic environment, as derived with the help of the fluctuation-dissipation relation, is given by ~\citep[][]{ermak1978brownian}
\begin{align}
    \eta_{\mathrm{eff}}^{\mathrm{hydro}}[u] \approx \gamma_h |u|,
\label{etahydro}
\end{align}
where $\gamma_h$ is a constant proportional to the solvent viscosity.
It is worth mentioning that, in Ref.~\citep[][]{ermak1978brownian}, no effective conservative forces are emerging from the hydrodynamic environment.
This means that the hydrodynamic scenario discussed in Ref.~\citep[][]{ermak1978brownian} behaves more similarly to our modified bilinear model than to the standard nonlinear model (which gives rise to $V_e[u]$, as we have seen in Eq.(\ref{ve})).

To model Eq.(\ref{etahydro}), we can simply choose
\begin{align}
    G(k,u) = k|u|,
\end{align}
since
\begin{align}
    \eta_\mathrm{eff}[u] 
    = \eta \int_0^\infty dk \ k^2 \  \frac{e^{-\frac{k}{k_0}}}{2k_0^3} \ k|u|
    = \gamma_m |u|,
    \label{etagammau}
\end{align}
with $\gamma_m = 3 \eta k_0$.

We have thus found a simpler way to map an environment-mediated dissipation of hydrodynamic nature into a fictitious bath consisting of a continuous set of harmonic oscillators.
By contrast, it is not clear how to derive Eqs.(\ref{etagammau}), and neither (\ref{etagauss}), from the standard nonlinear model, since the appropriate choice for the system-environment couplings as functions of the particles positions in the Lagrangian is not evident to us.

\section{Conclusions} \label{sec:conclusions}

In summary, we have revisited the standard bilinear and the standard nonlinear models for the dynamics of two Brownian particles in a collective environment.
In particular, we have addressed a controversy between these approaches, namely, whether distance-dependent nonlinear dissipation forces mediated by the environment, and affecting both the center of mass and the relative coordinate, should or not exist, and in which circumstances.

Our main result was the derivation of a nonlinear effective dissipation rate $\eta_{\mathrm{eff}}[u]$ departing from a bilinear Lagrangian.
Our method was based on the introduction of a distance-dependent nonlinear spectral function (response function) $\chi_k^{(ij)}(\omega,u)$, which settles a length scale to the dynamics of the Brownian particles (as also achieved with the standard nonlinear model, but not with the standard bilinear model).
This allows us to interpolate the dissipation forces as derived from the standard bilinear and the standard nonlinear models.
We have also shown that the nonlinear effective potential $V_e[u]$ obtained from the standard nonlinear model did not arise from our modified bilinear model.

As applications, we have discussed some consequences of tuning our distance-dependent spectral function.
For instance, a change from a trigonometric function to a Gaussian in $G(k,u)$ made it possible to avoid the anomalous diffusion presented in the original nonlinear model.
Also, we described hydrodynamic correlations between a pair of Brownian particles in a viscous fluid by means of our phenomenological distance-dependent spectral function.
These examples illustrate how our results represent a simple and versatile way to express diverse nonlinear dissipative forces in the dynamics of pairs of Brownian particles.

As a perspective, we would like to generalize our distance-dependent spectral function to a larger number of Brownian degrees of freedom in a common environment ($N$ particles in a three-dimensional space).
We believe it to be a feasible goal, given the pairwise character that typically underlies effective interactions.
This could allow us to characterize relaxation processes and entropy production in nonequilibrium dissipative many-body systems, across classical and quantum regimes~\citep[][]{landi2021irreversible}.
For instance, we could think of generalizing a recent study concerning entropy production of a single quantum Brownian particle ~\citep[][]{weiderpass2020neumann}.

\section*{Acknowledgements} \label{sec:acknowledgements}
EIG and RJSA ackowledge support from Coordena\c{c}\~{a}o de Aperfei\c{c}oamento de Pessoal de N\'{i}vel Superior - Brasil (CAPES) - Finance Code 001. DOSP ackowledge support by Brazilian funding agencies CNPq (Grant 307028/2019-4) and FAPESP (Grant No 2017/03727-0). DV was supported by the Serrapilheira Institute (Grant No. Serra-1912-32056). EIG, RJSA, DOSP and DV acknowledge support from the Instituto Nacional de Ci\^{e}ncia e Tecnologia de Informa\c{c}\~{a}o Qu\^{a}ntica, CNPq INCT- IQ (465469/2014-0), Brazil.

\appendix*
\appendix
\section{Two-time correlation functions and the fluctuation dissipation theorem}\label{appendix1:apendix1}

From the condition of thermal equilibrium, we have the following identities

\begin{align}
    \langle{\Tilde{R}_k(0)}\rangle &= 0, && \langle{\dot{R}_k(0)}\rangle = 0,\\
    \langle{\Tilde{R}_k(0)\dot{R}_{k'}(0)}\rangle &= 0, && \langle{\dot{R}_k(0)\Tilde{R}_{k'}(0)}\rangle = 0,\\
    \langle{\dot{R}_k(0)\dot{R}_{k'}(0)}\rangle &= \frac{k_B T}{m_k}\delta_{kk'}, && \langle{\Tilde{R}_k(0)\Tilde{R}_{k'}(0)}\rangle = \frac{k_B T}{m_k\omega_k^2}\delta_{kk'},
\end{align}
where $k_B$ is the Boltzmann constant.
As defined in the main text, the formal expressions for the fluctuating forces are given by
\begin{align}
    \begin{split}
        f_i(t) = -\sum_k C_k^{(i)}\left[\dot{R}_k(0)\frac{\sin{\omega_k t}}{\omega_k}+\Tilde{R}_k(0)\cos{\omega_k t}\right],
    \end{split}
\end{align}
where the displaced equilibrium positions of the oscillators (due to their couplings with the particles) are 
$\Tilde{R}_k(0) = R_k(0)+(C_k^{(i)}x_i(0)+C_k^{(j)}x_j(0))/(m_k \omega_k^2)$.
We have also defined
$f_q = (f_1+f_2)/2$, and 
$f_u = f_1 - f_2$.
With the above expressions at hands, we obtain the general form for the two-time correlation functions,
\begin{align}
    \langle f_{\alpha}(t)f_{\beta}(t') \rangle = 2 D_{\alpha \beta}(u) \delta(t-t'),
    \label{generalfluctuationdissipation}
\end{align}
where $D_{\alpha \beta}(u)$ is a type of diffusion coefficient having a different form according to the choice of forces we are dealing with.
To explicitly compute them, we apply the continuum limit in the same way we did in Sec.\ref{mainresult}.
We thus find Eqs.(\ref{D12}), (\ref{Du}), and (\ref{Dq}) as results.

For the sake of completeness, and also to highlight that Eq.(\ref{generalfluctuationdissipation}) is indeed a general form of the fluctuation-dissipation theorem, we recall the original version of the theorem below.
The Langevin equation for a free Brownian particle reads
\begin{align}
    m\dot{v}(t)= -\eta v(t)+F(t),
\end{align}
where $\eta$ is a friction coefficient, and $F(t)$ is a fluctuating force caused by collisions of the particle with the atoms of the surrounding fluid.
The fluctuating force fulfills
\begin{align}
    \langle{F(t)}\rangle=0, && \mbox{and} \ \langle{ F(t)F(t')}\rangle = 2D\delta(t-t'),
\end{align}
where $D$ can be seen as a measure of the strength of the fluctuating force. 
The delta function in time indicates that there is no correlation between impacts at any distinct time intervals.
The solution for the linear, first-order, and inhomogeneous differential equation reads~\citep[][]{zwanzig2001nonequilibrium}
\begin{align}
    v(t) = e^{-\eta t/m}v(0)+\int_{0}^{t}dt'e^{-\eta(t-t')/m} F(t')/m.
\end{align}
We can get the mean squared velocity and evaluate for long times, thus 
\begin{align}
    \langle{v^2(\infty)}\rangle = \frac{D}{\eta m}.
\end{align}
At thermal equilibrium, $\langle{v^2}\rangle_{eq} = k_B T/m$ (equipartition theorem), hence 
\begin{align}
    D = \eta k_B T.
\end{align}
This relates the strength $D$ of the random noise, or fluctuating force, to the magnitude $\eta$ of the friction, or dissipation rate, explaining why it is known as the fluctuation-dissipation theorem.
It expresses the balance between friction and noise that is required to have thermal equilibrium state at long times.

%
\end{document}